\documentstyle [12pt,graphicx,doublespace] {article}
\makeatletter
\renewcommand\section{\@startsection {section}{1}{\z@}%
                                   {-4ex \@plus -1ex \@minus -.2ex}%
                                   {4ex \@plus.2ex}%
                                   {\normalfont}}
\renewcommand\subsection{\@startsection{subsection}{2}{\z@}%
                                     {-4ex\@plus -1ex \@minus -.2ex}%
                                     {4ex \@plus .2ex}%
                                     {\normalfont}}
\renewcommand\subsubsection{\@startsection{subsubsection}{3}{\z@}%
                                     {-4ex\@plus -1ex \@minus -.2ex}%
                                     {4ex \@plus .2ex}%
                                     {\normalfont}}
\makeatother

\begin{document}
\textwidth=14cm
\textheight=21.5cm
\pagestyle{plain}
{\noindent\small revised: July 11, 2003}   % submit date
%--------------------- Title ----------------------------------------
\begin{singlespace}
\vspace*{0.6cm}\noindent
NUMERICAL INVESTIGATION OF A MESOSCOPIC VEHICULAR TRAFFIC FLOW MODEL
BASED ON A STOCHASTIC ACCELERATION PROCESS 
\vspace{24pt}\\
K.T. Waldeer 
\vspace{24pt}\\ 
Department of Transportation and Traffic \\
University of Applied Sciences Braunschweig/Wolfenb\"uttel\\
Karl-Scharfenberg-Str. 55, 38229 Salzgitter\\
Germany \\
Th.Waldeer@FH-Wolfenbuettel.de
%-------------------- abstract ---------------------------------------
\section{ABSTRACT}
In this paper a spatial homogeneous vehicular traffic flow
  model based on a stochastic master equation of Boltzmann type in the
  acceleration variable  
is solved numerically for a special driver interaction model.
The solution is done by a modified direct simulation Monte Carlo
  method (DSMC) well known in non equilibrium gas kinetic. The
  velocity and acceleration distribution functions in stochastic
  equilibrium, mean velocity, traffic density, ACN, velocity scattering
  and correlations between some of these variables and their car
  density dependences are discussed. 
\end{singlespace}
%---------------------------------------------------------------------
\section{INTRODUCTION}
Mesoscopic vehicular traffic flow models are based on stochastic
methods, mainly using master equations describing the time propagation
of a traffic state probability function either of single cars or car
clusters \cite{PH71,BKR94,KKW96,Hel97,SRM99}. Especially for single
car states Boltzmann like master equations are discussed in
literature. Therefore those models are often called kinetic
models. Mesoscopic models present the intermediate step between
microscopic single car behavior and macroscopic traffic flow. On the
one hand, like in gas kinetic the huge number of parameters in a
microscopic model can be reduced into mesoscopic interaction
probabilities, on the other hand, dynamic macroscopic flow equations
can be constructed from master equations using moment methods
\cite{Pav75,Phi77,Hel97}. Nearly all kinetic models assume that the
single car state consists of the velocity and space coordinate at any
time. The interaction is restricted to a leading car pair, where only
the following car changes its velocity using a stochastic jump
process. Like in gas kinetic this procedure only seems valid if the
mean durance of an interaction is much smaller than the mean time
between any two interactions, which only holds up to moderate
vehicular densities \cite{Gra58}. Single car measurements of the speed
and acceleration time development show that the durance of acceleration
changes are much shorter than the change of any other kinematic
variable \cite{Hoe72}. Therefore using an acceleration stochastic jump
process seems to be the natural choice in mesoscopic car
following. This idea is additional supported by the fact that the natural
driver control variable in an interaction is the acceleration.  \par
Based on these assumptions a new mesoscopic model for the single car
state probability density function in the space, velocity and
acceleration variables was constructed rigorously under Markov
conditions by the author reaching in an Enskog type master equation
\cite{Wal00}. To give a guess of its applicability simple
interaction functions were defined and first analytical results for a
homogeneous traffic flow were discussed. \par
In this article a more realistic driver interaction model is
constructed. Using this model, solving the master equation for a
homogeneous traffic flow, acceleration and velocity distributions and
important moments of the state probability density (i.e. mean
velocity, velocity scattering, acceleration scattering)  depending on
the car density in stochastic equilibrium are calculated.  \par
Section~\ref{s2} summarizes the model and its equations. In
section~\ref{s3} the interaction model is introduced. For solving the
master equation a numeric procedure borrowed from gas kinetic is used
and shortly described in section~\ref{s4}. In Section~\ref{s5} the
results are shown and discussed.  
%------------------------------------------------------------------------
\section{SUMMARY OF THE BASIC MODEL}
\label{s2}
In this section the basic homogenous traffic flow model equation
derived in \cite{Wal00} and its requirements are shortly summarized.  
At time $t$ the leading car has velocity $\bar{v}$ and acceleration
$\bar{a}$ and the following car has velocity $v$ and acceleration $a$
neglecting the spatial coordinates due to homogeneity. This pair of
cars is called a {\sl leading} car pair.  
The interaction between the cars in a given leading car pair is
assumed to be a Markov jump process in the acceleration variable $a$
of the following car. 
Therefore the time propagation of the car pair state probability
density fulfills the Feller-Kolmogorov equation. Introducing a
vehicular chaos ansatz \cite{Nel95, Wal00}, i.e. a leading car pair is only
distance correlated, and assuming that an interaction does not depend on the
acceleration of the leading car $\bar{a}$, because its value can not
be determined by the driver of the following car, the spatial
homogenous single car probability density $f$ is given by the solution of  
\begin{eqnarray}
\label{1}
&&{\partial f\over\partial t}+a{\partial f\over\partial v}=
\int_{\bar{v},a^\prime} d\bar{v}\, da^\prime\,
\tilde{f}(\bar{v},t)\cdot\nonumber \\ 
&&\quad\left\lbrace 
\Sigma(a|v,\bar{v},a^\prime, {\bf m}_f(t))f(v,a^\prime,t)
-\Sigma(a^\prime|v,\bar{v},a,{\bf m}_f(t))f(v,a,t)\right\rbrace\; .
\end{eqnarray}
$f(v,a,t)dv\,da$ is the probability to find a single vehicle at
velocities between $v$ and $v+dv$ and accelerations between $a$ and
$a+da$ at time $t$ and $\tilde{f}(v,t)=\int_a f(v,a,t)\, da$ is its
reduced density. 
The weighted interaction density $\Sigma$ includes the rate $Q$ of a
single interaction for a given car pair state $(v,a,\bar{v})$
and the interaction strength probability density $\sigma$ weighted by
a distance correlation function $D$, i.e.  
\begin{equation}
\label{2}
\Sigma(a|v,\bar{v},a^\prime,{\bf m}_f(t))=\int_{h_{\min}}^\infty
\sigma(a|h,v,\bar{v},a^\prime )Q(h,v,\bar{v},a^\prime )
D(h|v,a^\prime,{\bf m}_f(t))\, dh \; .
\end{equation}
Here $D$ is equal to the conditioned probability density of the
distance $h$ depending on the state of the following car and the
probability density $f$ or some moments of $f$ called moment
vector ${\bf m}_f$ at time $t$. In this paper, as described below,
the conditioned 
distance density  is chosen in such a way that the car density $K$ is
included into ${\bf m}_f$ as an constant, additional element. Note
that $D$ vanishes for $h$ lower 
than the minimal distance $ h_{\min}$ in a leading car pair at rest. A
further specification of these functions will be done in
section~\ref{s3}.\par  
Eq.~\ref{1} shows the typical structure of a stochastic rate equation.
 The total rate of change of $f$ at the state $(v,a)$ at time $t$ is 
equal to the difference of probability inflow in this state and
outflow out of this state. The vehicular chaos assumption, analog to
gas kinetic, leads in the typical Boltzmann like product form of $f$
on the right side. Note that in this paper a car following model
without any overtaking is discussed.  \par  
Additional to the equation an initial condition and boundary
conditions in $v$ must be added. 
The boundary conditions introduced here are described in \cite{Wal00}
and ensure that negative velocities $v<0$ are
impossible. In the case that a braking car reaches $v=0$, independent
on the state of its leading car, the driver changes the acceleration to
$a=0$. The interaction described in the next section is distance
oriented, and therefore does not describe traffic flow at low car
densities, i.e. the free flow regime. This regime is included by a
maximal velocity $w$ via a high speed boundary condition. An
accelerating car reaching $v=w$ changes the acceleration to $a=0$ to
hold this velocity, independent on the behavior of the leading
car. Because of this boundary condition velocity and acceleration
scattering do not occur in this regime in contrast to measurements
where slight scattering is always present. To relax this
unrealistic behavior, the boundary condition at $v=w$
must be changed into a relative velocity dependent interaction at high
velocities or large distances as
new unpublished developments on the model show. Further work on this
field is under way.\par
Because in this paper only the stochastic equilibrium is discussed,
the concrete form of the initial condition is not important. Here a
Gaussian velocity distribution with vanishing acceleration is
used. The simplest way to reach the stochastic equilibrium density
is to calculate the time propagation of a homogeneous traffic
flow, bearing in mind that in reality it remains not homogeneous during
this process. 
%----------------------------------------------------------------------------
\section{DESCRIPTION OF THE CAR PAIR INTERACTION} 
\label{s3}
In this section the interaction and distance correlation functions, defined in
Eq.~\ref{2} are concretized. Here a threshold or action point
interaction often described in literature for microscopic or
mesoscopic models \cite{Wie74,WK96} is used. For simplicity the
interaction is restricted to a single velocity dependent deterministic
threshold for all cars. Therefore the driving behavior must be
conservative to avoid accidents.  
%........................................................................
\subsection{The Interaction Rate}
In the action point model, interactions only occur on velocity
dependent distance thresholds, redefining the interaction rate $Q$
compared to those described in \cite{Wal00}.
In a given leading car pair the following car with velocity $v$
interacts, i.e. changes its acceleration state, if the distance $h$ to
the leading car is equal to an action point or threshold distance
$H(v)$. So $h=h(t)$ and $H(t)=H(v(t))$ are both functions of time
$t$. Staying at time $t$ shortly before the threshold,
i.e. $h(t)-H(t)\rightarrow 0^-$ or $h(t)-H(t)\rightarrow 0^+$,
assuming an interaction at time $t+\tau$ , where $\tau$ is
infinitesimal small,    
\begin{eqnarray}
\label{5}
H(t+\tau)&=&H(t)+H^\prime(v)\cdot a\tau\quad\mbox{with}\quad
H^\prime(v)={dH\over dv}\; ,\nonumber \\
h(t+\tau)&=&h(t)+(\bar{v}-v)\tau 
\end{eqnarray}
are the threshold and distance values directly after this interaction
up to the first order in $\tau$. An acceleration interaction $a> 0$ is
given by $h(t)-H(t)\,<\, 0$ and $h(t+\tau)-H(t+\tau) \ge 0$, where a
de acceleration interaction, $a< 0$, is given by $h(t)-H(t)\, >\, 0$ and
$h(t+\tau)-H(t+\tau)\le 0$. Inserting this into Eqs.~\ref{5}, lead to necessary
conditions for acceleration $\bar{v}-v-H^\prime(v)\cdot a\, >\, 0$ and
de acceleration $\bar{v}-v-H^\prime(v)\cdot a \le 0$. Because an
interaction occurs in a deterministic way when the distance between the two
cars passes the threshold, the probability of the acceleration
interaction $P^+(\tau)$ and the de acceleration interaction
$P^-(\tau)$ are given by 
\begin{eqnarray}
\label{6}
P_+(\tau)&=&\Theta(h(t+\tau)-H(t+\tau))\Theta(\bar{v}-v-H^\prime(v)\cdot
a)\;
,\nonumber \\
P^-(\tau)&=&\Theta(H(t+\tau)-h(t+\tau))\Theta(-(\bar{v}-v-H^\prime(v)\cdot
a))\; ,
\end{eqnarray}
where $\Theta(x)$ is the Heaviside unit step function. The probability
to have an interaction until $\tau$ now is $P^+(\tau ) +P^-(\tau )$,
and its derivative at $\tau=0$ with Eqs.~\ref{5} is the interaction
rate 
\begin{equation}
\label{7}
Q(h,v,\bar{v},a,\bar{a})=|\bar{v}-v-H^\prime(v)\cdot a|\delta(h-H(v))\; ,
\end{equation}
where $\delta(x)$ is the Dirac distribution. A linear velocity
dependence   
\begin{equation}
\label{8a}
H(v)=h_{\min}+\alpha v
\end{equation}
with some given constant $\alpha$ specified below
is used, bearing in mind that this is a conservative driver behavior
ansatz at high velocities \cite{Wie74}. 
%........................................................................
\subsection{The Interaction Strength}  
The strength of an interaction is described by the probability density
of an acceleration or de acceleration change in analogy to the
differential cross section in gas kinetic. Because only a single
interaction threshold is used especially in the de acceleration
process one has to ensure $h \ge h_{\min}$ to avoid accidents. This is
only possible by defining the de acceleration change $a^-$ as the
total braking value of the following vehicle with given velocity $v$
until it stops in the distance $H(v)-h_{\min}+\bar{h}$, where $\bar{h}$ is
the minimum braking distance of the leading vehicle at velocity
$\bar{v}$ with maximum braking value $\bar{a}=a_{\min}< 0$:  
\begin{equation}
\label{8}
a^-={-v^2\over 2(H(v)-h_{\min}+h^*)}\quad \mbox{with}\quad
\bar{h}={\bar{v}^2\over 2 |a_{\min}|}\; . 
\end{equation}
The acceleration change $a^+$ depends strongly on the actual velocity
$v$ of the car. Where at very low velocities the acceleration value
increases, at higher velocities it decreases and vanishes near by the
maximum velocity $w$ \cite{CH80}. This is modeled using a linear increasing
function from $a^+=a_0\, >\, 0$ at $v=0$ to $a^+=a_{\max}\, >\, 0$ at
$v=v_m\ll w$ followed by a linear decreasing function starting a
$v=v_m$ and reaching $a^+=0$ at $v=w$, which is a simple caricature of
measured driver behavior: 
\begin{equation}
\label{9}
a^+=\left\{
\begin{array}{ll}
&{w-v\over T}\;,\quad v\ge v_m\; ,\\
&a_0+{a_{\max}-a_0\over v_m}v\quad\mbox{with} \quad a_{\max}={w-v_m\over
  T}\;,\quad v\, <\, v_m \; .
\end{array}
\right.
\end{equation}
Individual driver behavior shows some scattering in the acceleration
strength. Here for simplicity and to minimize the number of parameters
this scattering is avoided. Therefore the interaction strength density
at $h=H(v)$ is given by 
\begin{equation}
\label{10}
\sigma(a|H(v),v,\bar{v},a^\prime,\bar{a})=\left\{
\begin{array}{ll}
&\delta(a-a^+)\; ,\quad \bar{v}-v-H^\prime(v)\cdot a^\prime\, >\; 0\; ,\\
&\delta(a-a^-)\; ,\quad \bar{v}-v-H^\prime(v)\cdot a^\prime \le 0\; ,
\end{array}
\right.
\end{equation}
where the decision of accelerating or de accelerating depends on the
sign of $\bar{v}-v-H^\prime(v)\cdot a^\prime$ as is discussed above. \par
It is well known that driver behavior in acceleration and de
acceleration
depends on the relative velocity between the leading and the following
car \cite{ER73, Leu88}. This behavior homogenizes the traffic
flow. Comparing the simulation results (see below) to those found in
literature show a qualitative agreement for velocity dependent
quantities.  In contrast to this the acceleration distribution shows
significant deviations mainly reducible to the lack of relative
velocity dependence of the acceleration change in the model \cite{Wit96}.   
To integrate such an effect into the interaction, a second, more
distant threshold must be used additionally to the profile used
here. On this threshold a relative velocity dependent acceleration
change can be defined. Further work to the influence on the resulting
quantities and comparison to measured data is underway by the author.
%........................................................................
\subsection{The Spatial Correlation}
The spatial correlation is given by the following driver dependent
distance probability density $D(h|a,v,{\bf m}_f)$. For the single
threshold interaction model the sign of the acceleration $a$ of the
following car is equal to the sign of $h-H(v)$. Restricting to the
case where the velocity dependence is only given via $H(v)$ the
following ansatz can be used for acceleration and de acceleration 
\begin{eqnarray}
\label{11}
D(h|a\, <\, 0,v,{\bf m}_f)&=&D(h|\, h\le H(v),{\bf m}_f)={\tilde{D}(h|{\bf
    m}_f)\over\int_{h_{\min}}^{H(v)} \tilde{D}(h|{\bf m}_f) dh}\; ,
    \nonumber \\
D(h|\, a\ge 0,v,{\bf m}_f)&=&D(h|h\, >\, H(v),{\bf m}_f)={\tilde{D}(h|{\bf
    m}_f)\over\int_{H(v)}^\infty \tilde{D}(h|{\bf m}_f) dh}\; . 
\end{eqnarray}
The second equality is given by the conditioned probability standard
theorem \cite{Pap65}. Here $\tilde{D}(h|{\bf m}_f)$ is the driver
unconditioned distance probability density with parameters depending
on moments of the unknown $f$. By inserting Eq.~\ref{11} into Eq.~13
of paper \cite{Wal00} it is shown that this equation is
fulfilled identically. Therefore the car density $K$ is a free parameter
of the problem, which in the case of equilibrium can be chosen as
time-constant. It is included into ${\bf m}_f$. 
Normally distance measurements are done
locally resulting in time headway distributions, which are often
approximated by gamma densities \cite{May90}. Therefore in this
article a gamma density is also used neglecting possible differences
between time headway density and distance density      
\begin{equation}
\label{13}
\tilde{D}(h|r,\lambda)={\lambda\over\Gamma(r)}(\lambda(h-h_{\min})^{r-1}e^{-\lambda(h-h_{\min})}\Theta(h-h_{\min})\; 
, 
\end{equation}
where $\Gamma (x)$ is the standard gamma function \cite{AS72}.
Depending on the parameter values $\lambda ({\bf m}_f)\, >\, 0$ and
$r({\bf m}_f)\, >\, 1$ a wide spread of shapes can be constructed
including Gaussian, exponential and in the limit case delta
densities. These parameters are specified by the following
conditions. The mean distance, i.e. the mean value of $\tilde{D}$, is
equal to the inverse of the car density $K$ and the scattering of
$\tilde{D}$ here is modeled to be proportional to the mean velocity of
all cars $V$ (with constant $\beta$). The second ansatz seems to be
feasible, because with increasing mean velocity and therefore
decreasing car density the distance scattering increases also. These
two conditions together with Eq.~\ref{13} result in
$r=(K^{-1}-h_{\min})^2/(\beta V)^2$ and
$\lambda=(K^{-1}-h_{\min})/(\beta V)^2$. So the moment vector
dependence ${\bf m}_f$ is reduced to the mean velocity $V$ and the car
density $K$. Inserting Eq.~\ref{13} into \ref{11},
bearing in mind that due to Eq.~\ref{7} and the interaction
Eq.~\ref{2} only the value at $h=H(v)$ is needed, result in 
\begin{eqnarray}
\label{14}
D(H(v)|a\ge 0,v,K,V)&=&{(\lambda(H(v)-h_{\min}))^r
  e^{-\lambda(H(v)-h_{\min})}\over
  (H(v)-h_{\min})(\Gamma(r)-\gamma(r,\lambda(H(v)-h_{\min})))}\;
,\nonumber \\ 
D(H(v)|a < 0,v,K,V)&=&{(\lambda(H(v)-h_{\min}))^r
  e^{-\lambda(H(v)-h_{\min})}\over 
  (H(v)-h_{\min})\gamma(r,\lambda(H(v)-h_{\min}))}\; ,
\end{eqnarray}
where $\gamma(r,x)$ is the incomplete gamma function \cite{AS72}. So
the weighted interaction density, Eq.~\ref{2}, is totally specified by
Eqs.~\ref{7}, \ref{10} and \ref{14}. 
%----------------------------------------------------------------------------
\section{NUMERICAL REMARKS} 
\label{s4}
The solution of the master equation together with the interaction
functions specified is only possible numerically. Like in gas kinetic
this type of integro-differential equation is hard to implement on a
computer system using standard integration and differentiation
approximation rules. Nevertheless, there are stochastic solution
methods called 
{\sl direct simulation Monte Carlo} (DSMC) well established in non
equilibrium gas kinetic for the Boltzmann equation, which after some
modifications can be used following the ideas of Nanbu, 1981
\cite{Nan83}. Only a short description will be given, a detailed
analysis is published elsewhere \cite{Wal02,Wal03}.\par 
In DSMC methods the probability density $f$ is approximated by a
discrete measure, here a number $N$ of stochastic cars. Each car has a
velocity $v$ and an acceleration $a$. The simulation procedure is
split  into two parts over a discrete time interval $\Delta t$ small
against all characteristic timescales of the process. In the first
part, all cars are accelerated or de accelerated without any
interaction. In the second part, for each car there is done a decision,
whether it interacts with an other one due to the total interaction probability
$\Delta t\cdot\int_a\Sigma(a|v,\bar{v},a^\prime,{\bf m}_f)\,da$.  Any interaction is constructed by
changing the acceleration value of the following car using
Eq.~\ref{2}. Then again all cars change their velocities due to their
actual acceleration values. The simulation results in this paper are
based on $N=100$ cars, which are followed in time until the stochastic
equilibrium  density $f_{\mathrm eq}(v,a)$ is reached. The process is repeated
100 times to reduce numerical scattering. In each run the simulation
stops, when changes in the distribution are negligible and the mean
acceleration nearly vanishes.  
%----------------------------------------------------------------------------
\section{DISCUSSION OF THE RESULTS} 
\label{s5}
In this section the velocity and acceleration distribution together
with some of their moments like mean values, scattering, skewness and
correlations depending on the car density $K$ in stochastic
equilibrium are discussed. To reduce the number of parameters some
scaling is introduced. There are two interaction independent
parameters $h_{\min}$ and $w$ in the model, which together with the
acceleration scale $w^2/h_{\min}$ are used in the discussion. In this
section scaled variables and functions are marked by a prime. The
numerical solution is done using the following parameter values:
$\alpha^\prime=10$, $a_{\min}^\prime=-0.013$, $v_m^\prime=0.125$,
$T^\prime=86$ and $\beta^\prime=0.57$, which can be applied for
typical values of $h_{\min}\approx $7m and $w\approx $40m/s. So for
example the value of $\alpha$ is defined in such a way that the
threshold is approximately equal to half of the velocity in units of
km/h.\par 
Figs.~\ref{f1} and \ref{f2} show the typical fundamental diagram for
the mean velocity  
\begin{equation}
\label{15}
V=E[v]=\int_{v,a}v\,f_{\mathrm eq}(v,a)\,dv\,da
\end{equation}
and traffic density $q=V\cdot K$ depending on the car density
$K$. The early decrease of the mean velocity is due to the linear
threshold behavior, which at high velocities is a conservative driving
approximation.\par 
Defining the scattering of a variable $x$ as
$\sigma_x^2=E[(x-E[x])^2]$ using definition Eq.~\ref{15} for $E[x]$
analogously, figs.~\ref{f3} and \ref{f4} show the car density dependence
of the velocity and acceleration scattering resp.. Especially
the acceleration scattering, often called acceleration noise (ACN),
shows the same behavior as measurements predict \cite{Win79}. There at
very low car densities, in the free flow region, only a small amount
of ACN occurs, which vanishes for $K\rightarrow 0$ due to the boundary
condition at $v=w$. As mentioned above the free flow part is in
contrast to experimental results, where typical values of 0.3m/s$^2$ occur.
With increasing car density, in the area of partly constrained
traffic, there is a huge increase of ACN. At high densities -- the so
called constrained traffic area -- ACN decreases strongly, because all
driver are obliged to have nearly the same behavior. The
velocity scattering shows the same shape. Therefore a positive
correlation between both moments can be expected. This is shown in
fig.~\ref{f5} using the definition of centralized moments
${\mathrm Corr}(x^n,y^m)=E[(x-E[x])^n(y-E[y])^m]/(\sigma_x^n\sigma_y^m)$
for any two random variables $x$ and $y$. With increasing car
density the correlation ${\mathrm Corr}(v^2,a^2)$ reaches a
maximum, similar to that of the velocity and acceleration scattering.   
At higher car densities the decrease of scattering is shifted between both
quantities resulting in a slight decrease in the correlation.
In contrast to the scattering correlation in fig.~\ref{f5}, the correlation between mean
velocity and ACN, fig.~\ref{f6}, is
negligible as is obtained for simple interaction models introduced in
\cite{Wal00}. \par  
The skewness $S=E[(x-E[x])^3]/\sigma_x^3$ is a measure for the symmetry of
a distribution. Here fig.~\ref{f7} shows the skewness of the velocity
distribution. At small car densities there
is a strong increase from large negative values to zero, which re give
the transition from a delta peaked velocity density at $v=w$, over a
left skewed distribution in the transition area, until a symmetric
distribution at high car densities, where the boundary condition at
$v=w$ is not significant any longer. \par
In fig.~\ref{f8} the density dependence of the equilibrium mean
interaction rate defined by 
\begin{equation}
\label{16}
\nu =  \int_{\bar{v},v,a,a^\prime}
\Sigma(a|v,\bar{v},a^\prime,{\bf m}_f)f_{\mathrm eq}(v,a^\prime)
\tilde{f}_{\mathrm eq}(\bar{v})\, d\bar{v}\, dv\, da\,da^\prime 
\end{equation}
is plotted. As expected the interaction rate increases with increasing
car density 
starting at $K^\prime\approx 0.1$, where the partly
constraint traffic flow begins. Note that there is no decrease at high
car 
densities, because even when all cars are stopped in a traffic jam the
drivers want to accelerate, which produce a lot of small valued
interactions. \par  
Fig.~\ref{f9} shows velocity distributions for different $K$ values,
which are more or less Gaussian like, as is well known from 
measurements \cite{Hel97}. The analog acceleration distributions are
shown in fig.~\ref{f10}. They are strongly bimodal reflecting the
interaction strength density Eq.~\ref{10}, used in this
simulation. Note that measured distributions are much broader in zero
direction \cite{Wit96}. This difference is possibly due to the
single threshold interaction model used here as mentioned above. 
%---------------------------------------------------------------------
\section{CONCLUSIONS}
\begin{itemize}
\item[(1)]
A spatial homogeneous vehicular traffic flow model based on a
stochastic master equation of Boltzmann type in the acceleration
variable is
applied to a concrete interaction model. For solution a modified DSMC
method well known in non equilibrium gas kinetic is used. The velocity
and acceleration distribution functions in stochastic equilibrium and
some of their moments and correlations are discussed.  
\item[(2)]
The fundamental diagrams, i.e. the car density dependence of the mean
velocity and the traffic density, are in qualitative agreement with
other models or 
measurements. Calculated ACN shows the same qualitative behavior as
experiments predict. There is a huge increase of ACN at the transition
from free to partly constraint traffic. The velocity
scattering is positive correlated to the ACN.   
\item[(3)]
There is no significant skewness in the velocity distributions at
higher car densities. The velocity distributions show a symmetric and
more or less Gaussian distributed behavior. There is only a very small
correlation between the mean velocity and the ACN in a traffic
flow. Acceleration distributions are bimodal and strongly peaked in
contrast to experimental work. Here additional investigations
especially the introduction of a relative velocity dependent
acceleration strength in an interaction have to take place. 
\end{itemize}
%-------------------- Bibliography -----------------------------------
\begin{singlespace}

\end{singlespace}
%------------------------ Figure Captions ----------------------------------
\newpage
\begin{singlespace}
\section*{FIGURE CAPTIONS}
\newcounter{fig}
\begin{list}{Fig. \arabic{fig}:}{\usecounter{fig}}
\item\label{f1}
{\small Scaled mean velocity as a function of the scaled car density.}
\item\label{f2}
{\small Scaled traffic density as a function of the scaled car density.}
\item\label{f3}
{\small Scaled velocity scattering as a function of the scaled car density.} 
\item\label{f4}
{\small Scaled ACN as a function of the scaled car density.}
\item\label{f5}
{\small Scattering correlation between $\sigma_v$ and
  ACN as a function of the scaled car density.}  
\item\label{f6}
{\small Correlation between mean velocity $V$ and ACN as a
  function of the scaled car density. 
} 
\item\label{f7}
{\small Skewness of the velocity distribution as a function of
  the scaled car density.} 
\item\label{f8}
{\small Scaled mean interaction rate as a function of
  the scaled car density.} 
\item\label{f9}
{\small Velocity distributions for different values of
  the scaled car density.}   
\item\label{f10}
{\small Acceleration distributions for different values
  of the scaled car density.}   
\end{list}
\end{singlespace}
%-----------------------------------------------------------------------
\newpage
\begin{figure}[t]
\centering
\includegraphics[angle=90,height=20cm]{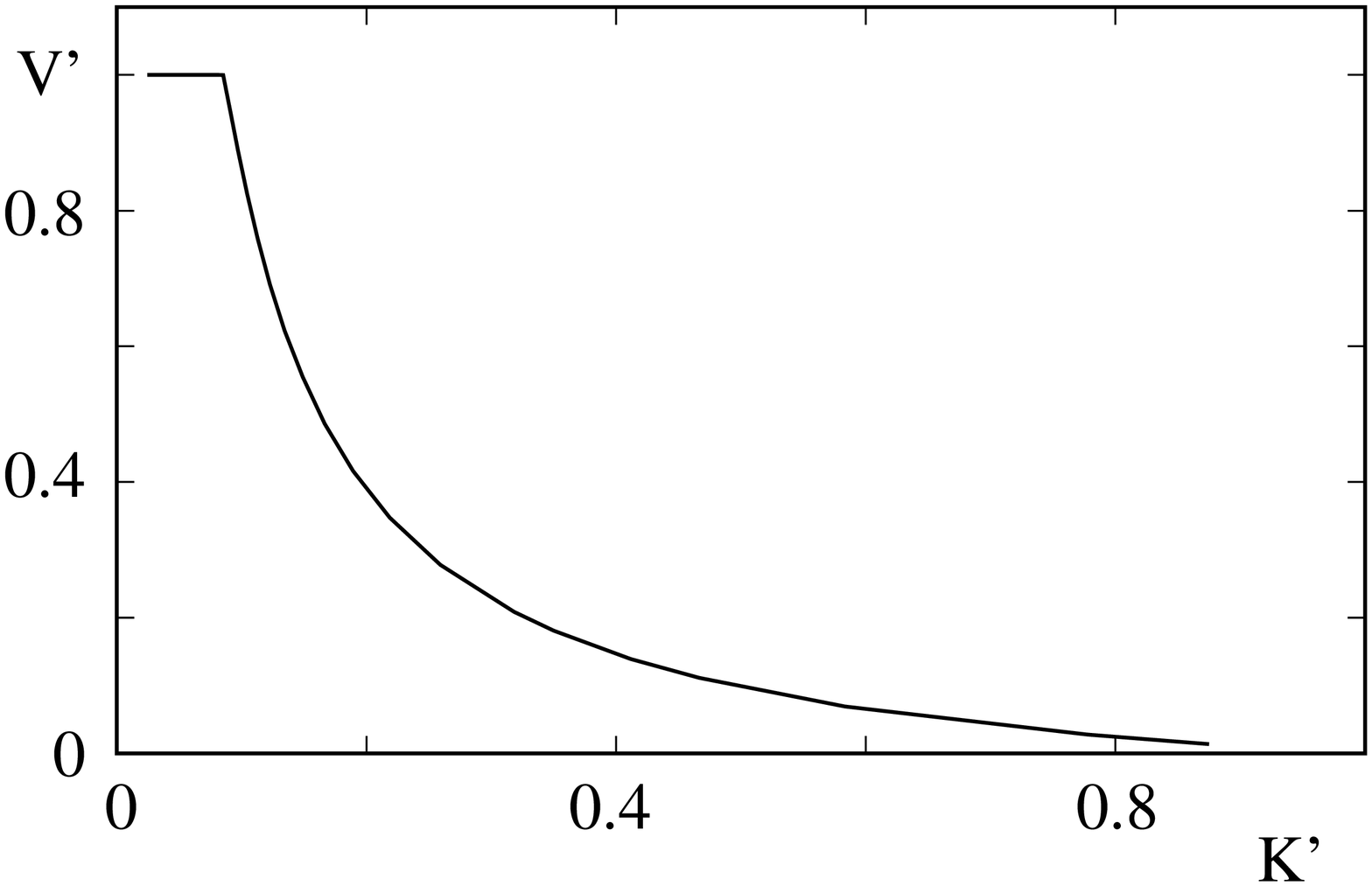}
\caption{}
\end{figure}
\begin{figure}[t]
\centering
\includegraphics[angle=90,height=20cm]{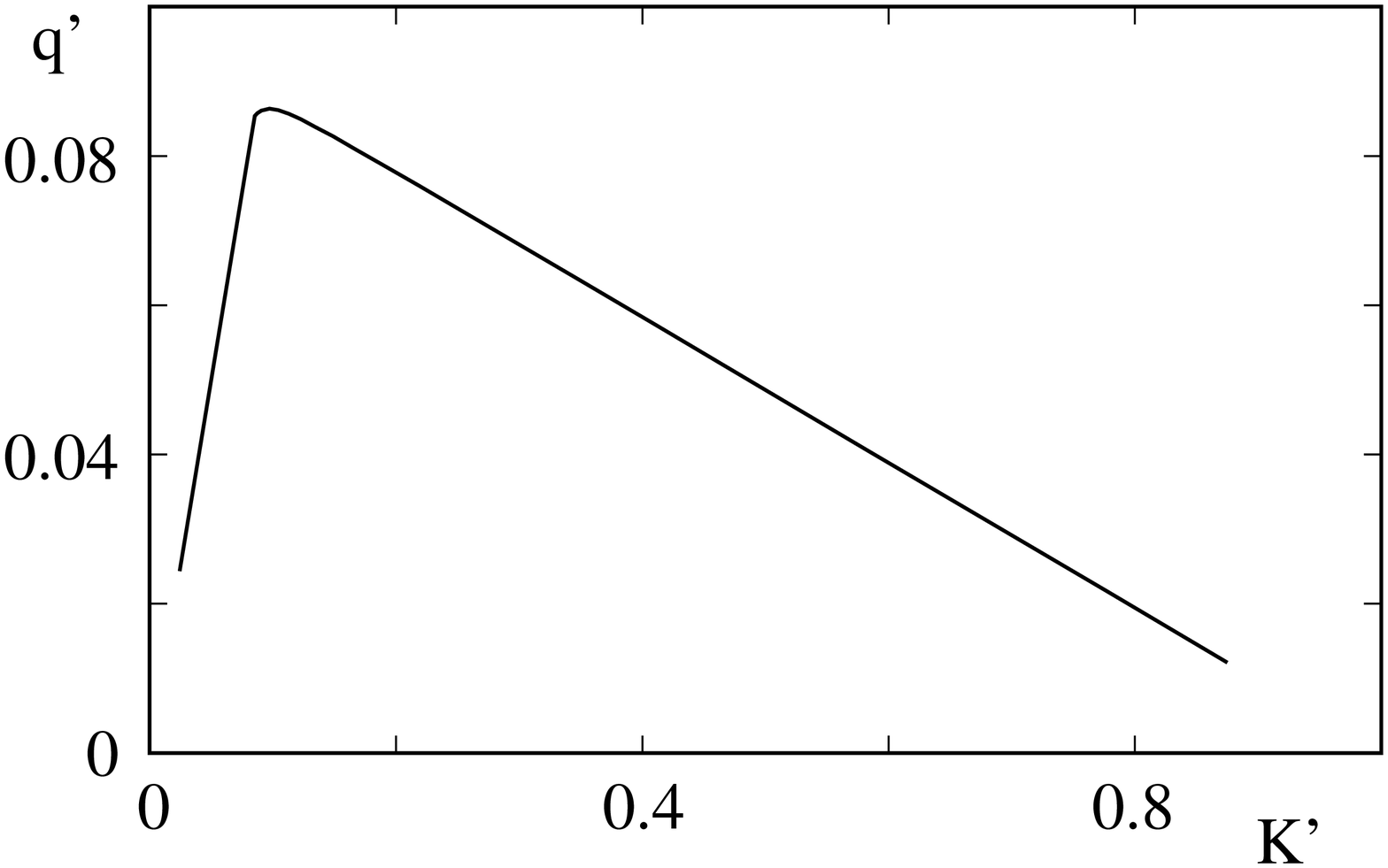}
\caption{}
\end{figure}
\begin{figure}[t]
\centering
\includegraphics[angle=90,height=20cm]{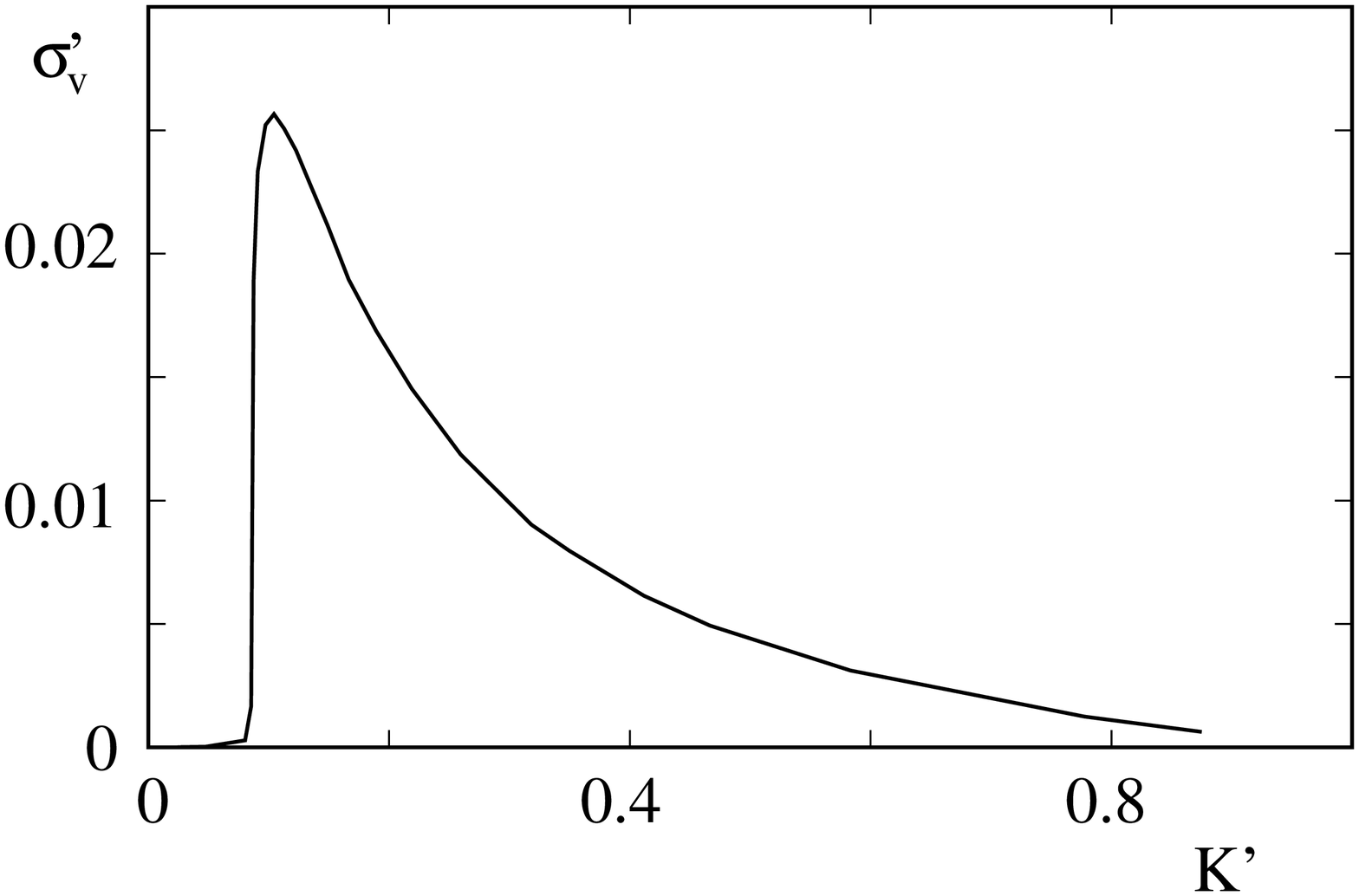}
\caption{}
\end{figure}
\begin{figure}[t]
\centering
\includegraphics[angle=90,height=20cm]{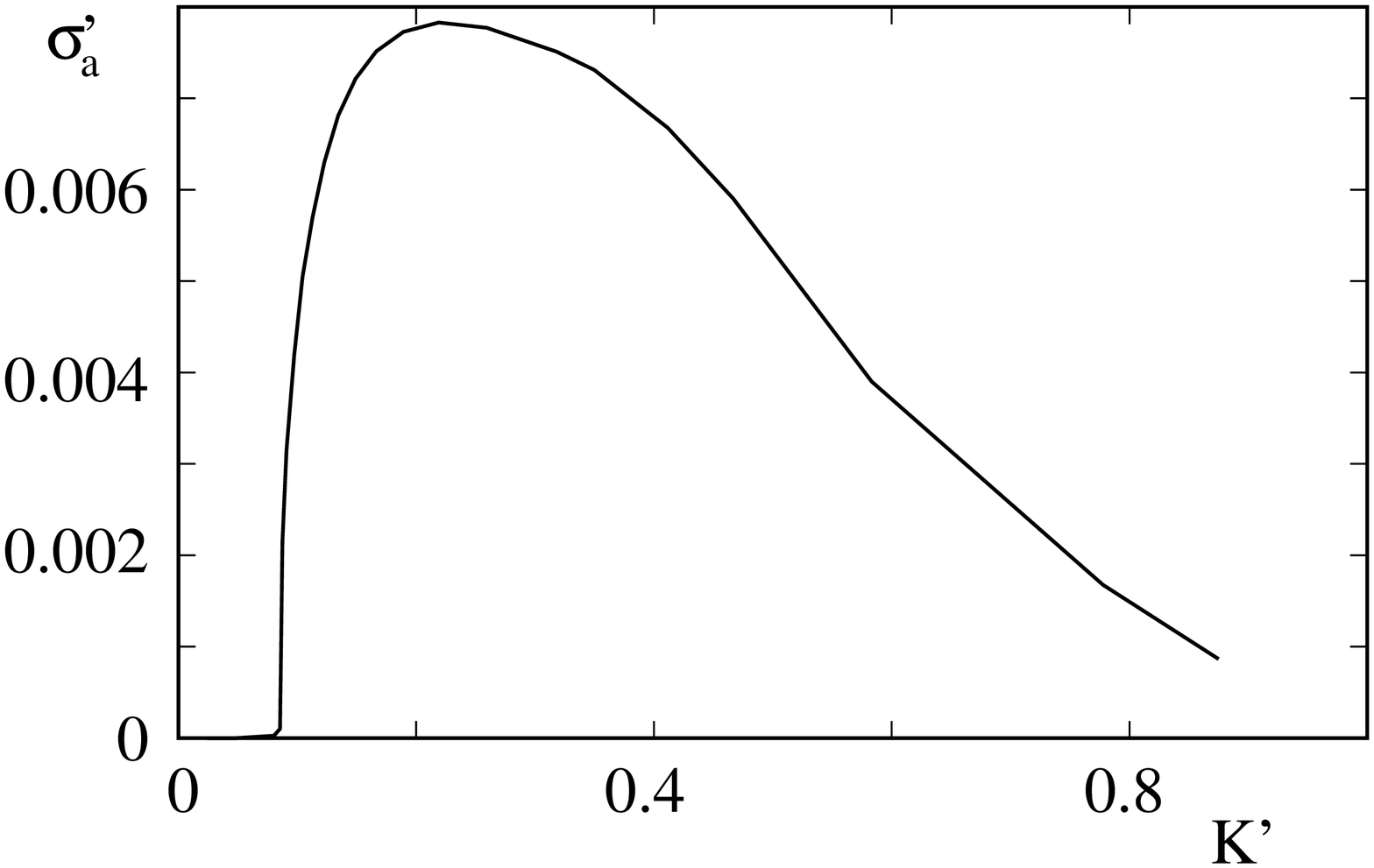}
\caption{}
\end{figure}
\begin{figure}[t]
\centering
\includegraphics[angle=90,height=20cm]{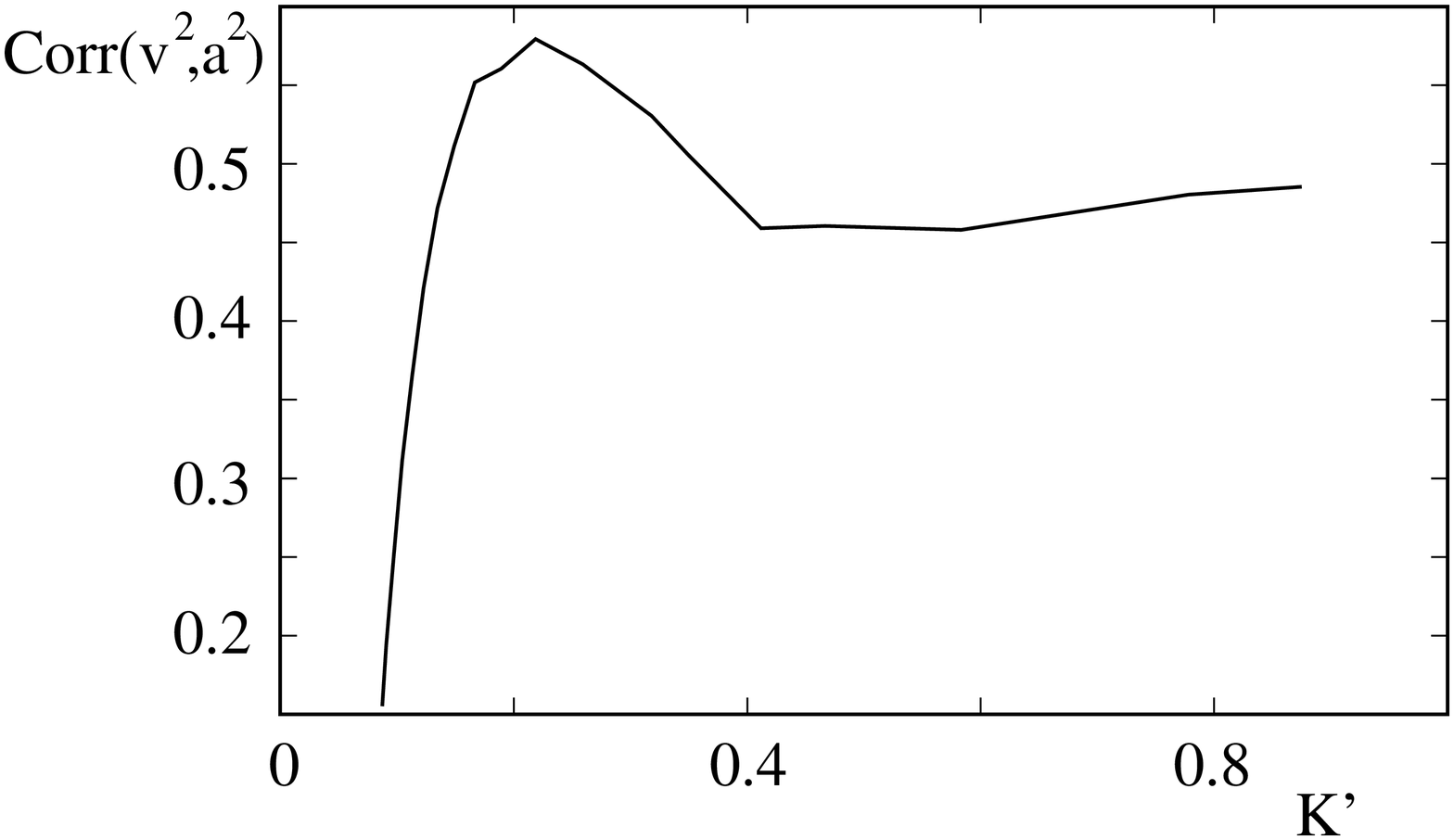}
\caption{}
\end{figure}
\begin{figure}[t]
\centering
\includegraphics[angle=90,height=20cm]{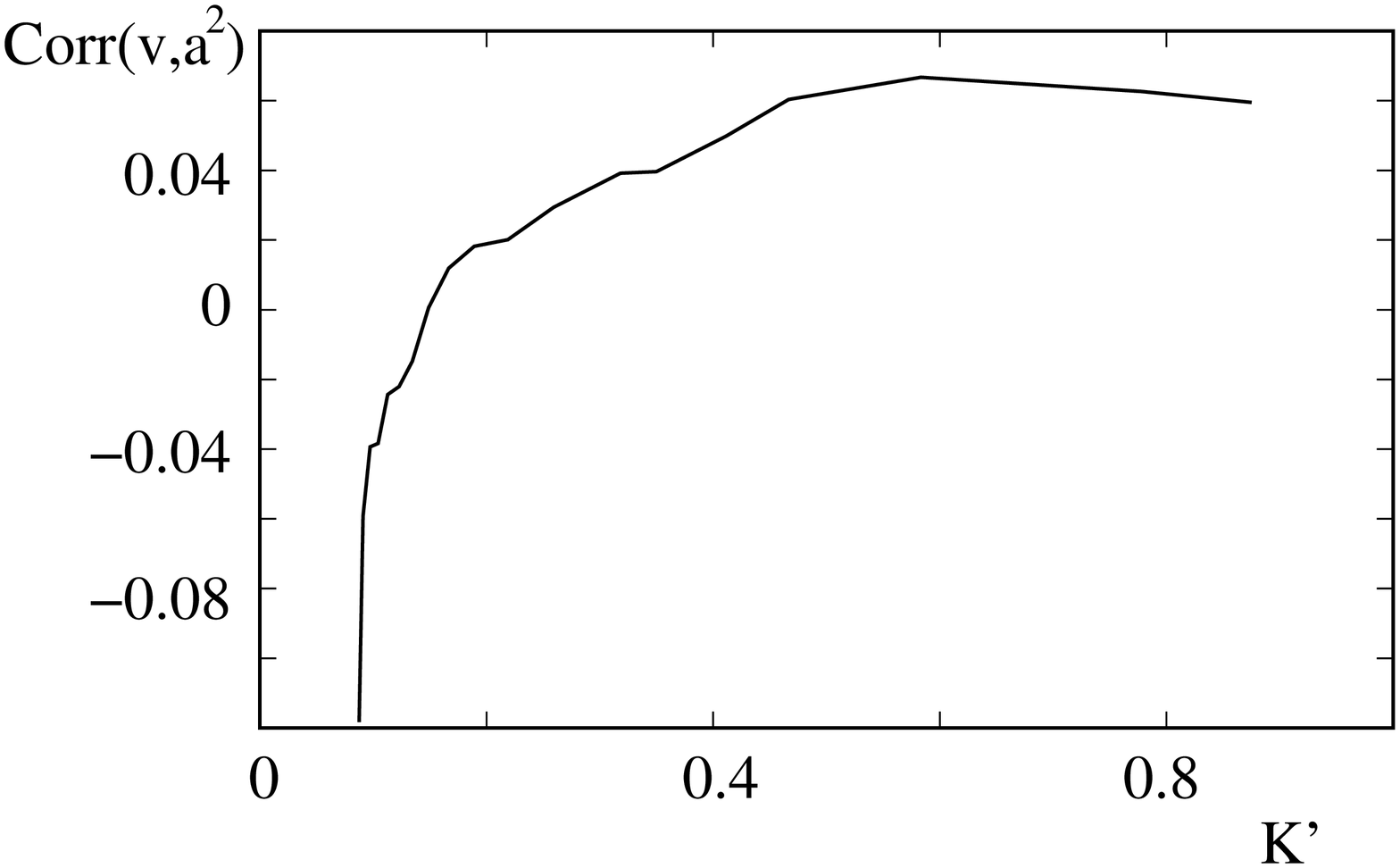}
\caption{}
\end{figure}
\begin{figure}[t]
\centering
\includegraphics[angle=90,height=20cm]{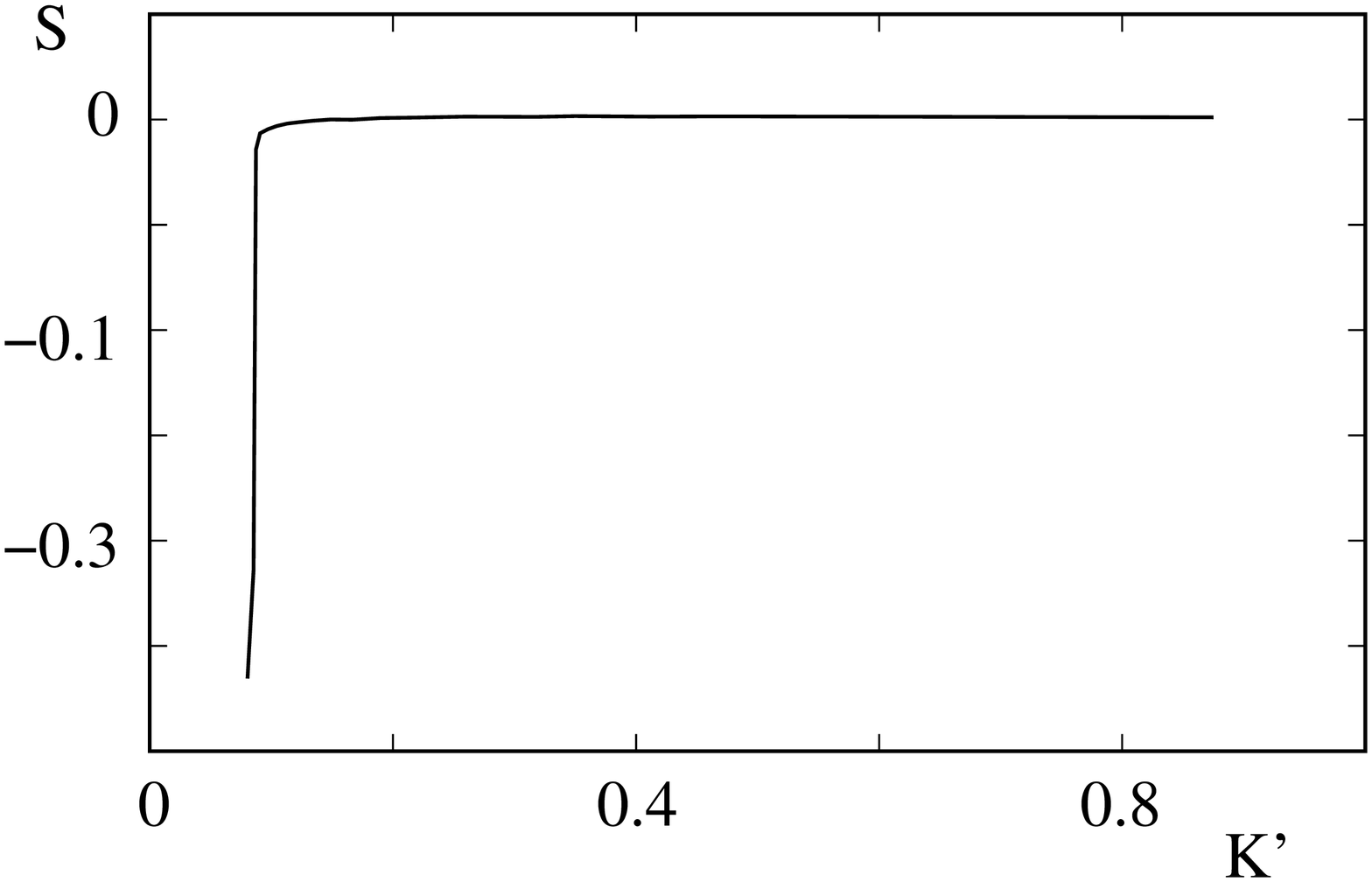}
\caption{}
\end{figure}
\begin{figure}[t]
\centering
\includegraphics[angle=90,height=20cm]{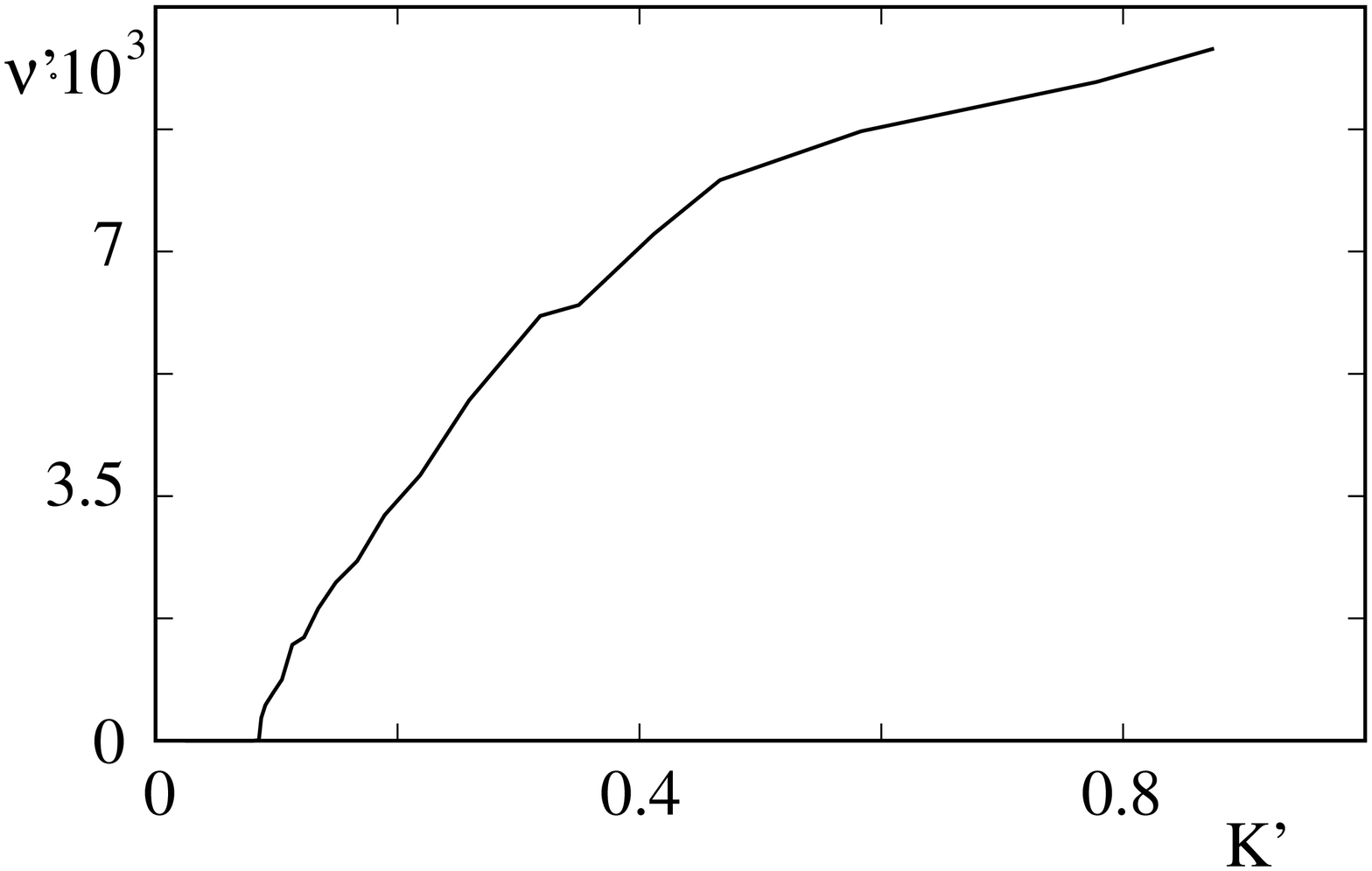}
\caption{}
\end{figure}
\begin{figure}[t]
\centering
\includegraphics[angle=90,height=20cm]{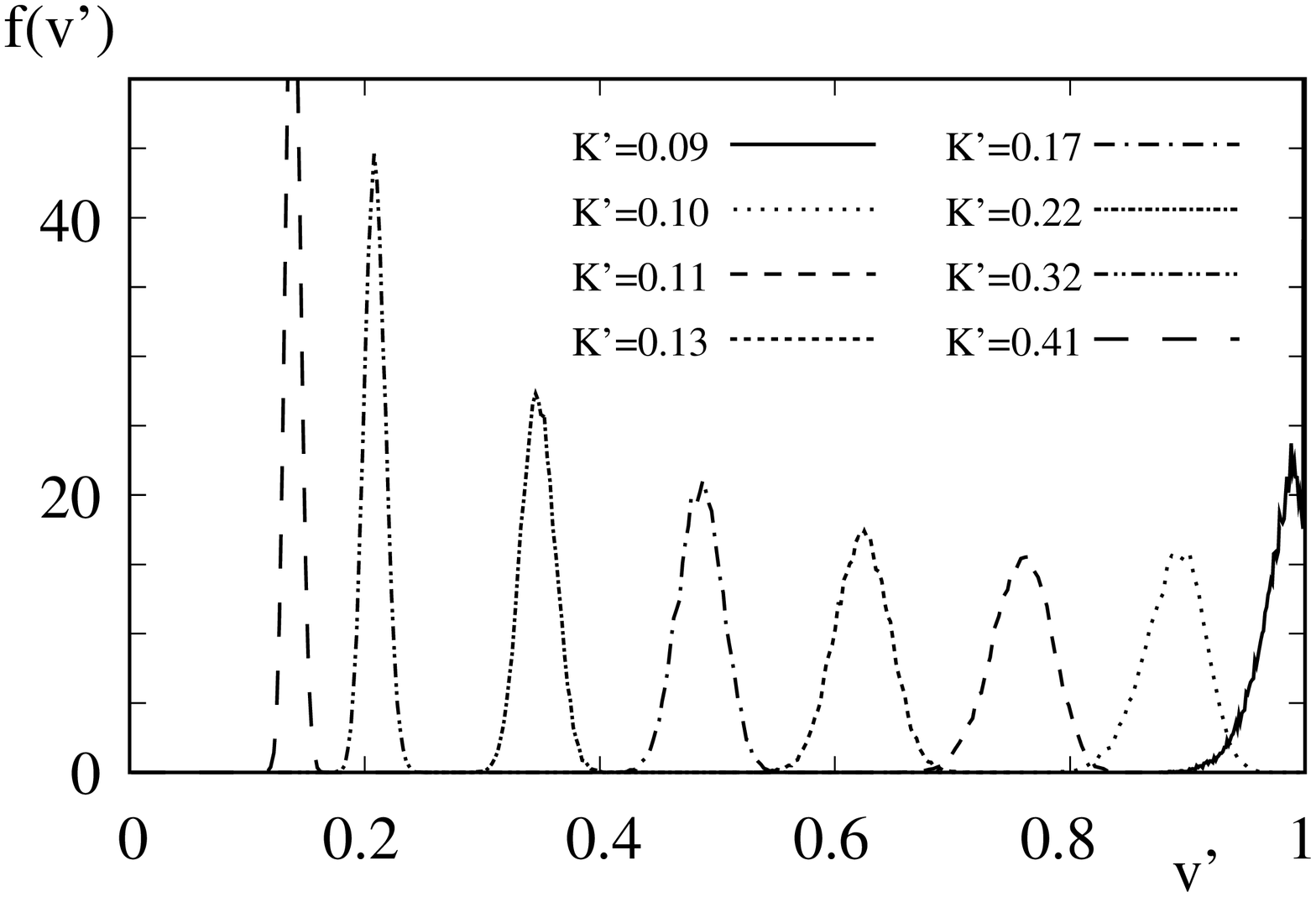}
\caption{}
\end{figure}
\begin{figure}[t]
\centering
\includegraphics{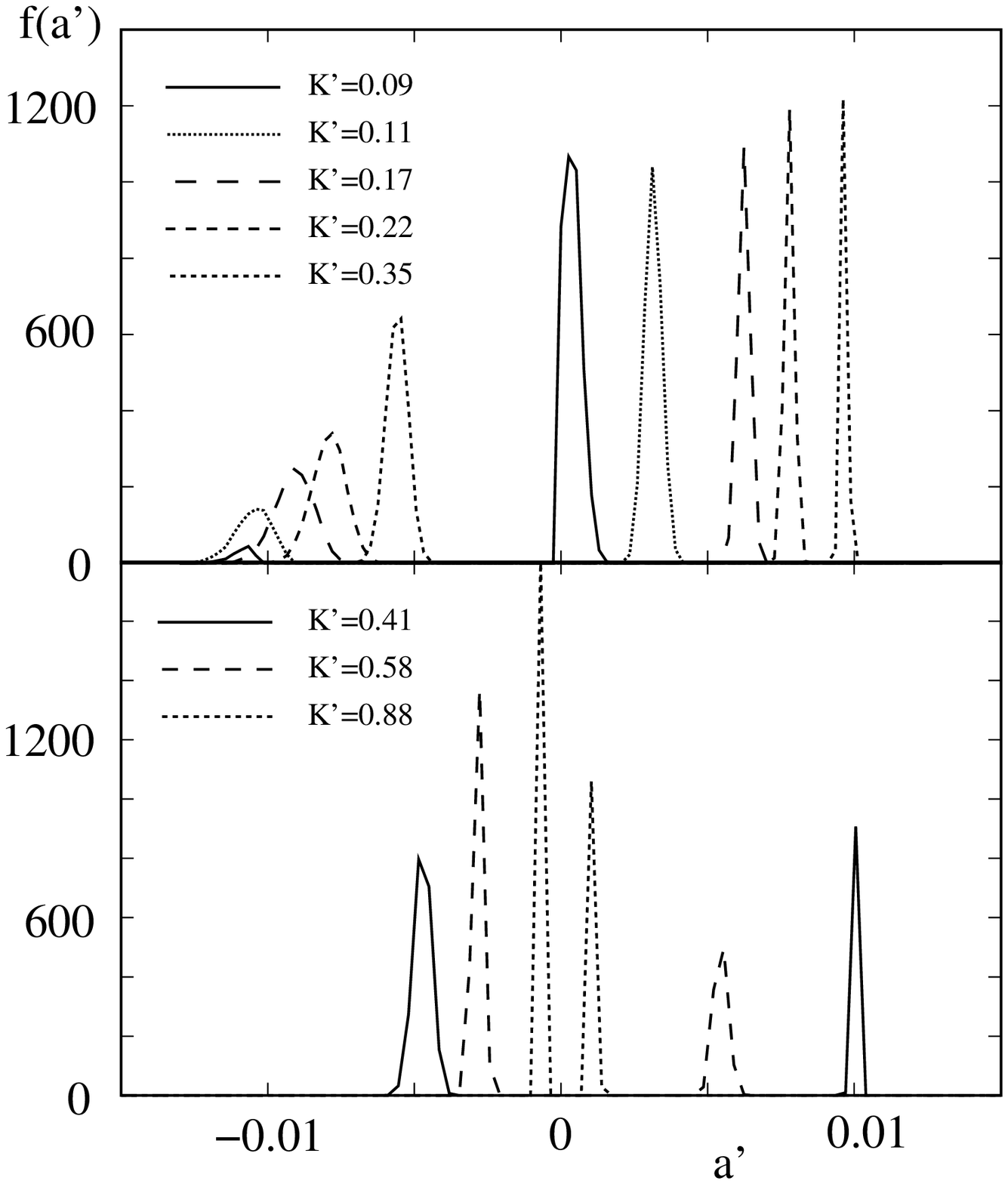}
\caption{}
\end{figure}
%---------------------------------------------------------------------
\end {document}